\documentclass[a4paper,12pt]{article}
\pdfoutput=1    

\usepackage[]{hyperref} 

\usepackage[utf8]{inputenc}
\usepackage{amsmath}
\usepackage{amsthm}
\usepackage{amssymb}
\usepackage{mathrsfs}
\usepackage{amsfonts}
\usepackage{bbold}

\usepackage{algorithm}
\usepackage{algorithmicx}
\usepackage{algpseudocode}
\usepackage{graphicx}
\usepackage{xifthen}
\usepackage{rotating}
\usepackage{xcolor}
\usepackage{subfig} 
\usepackage{psfrag}
\usepackage{amsopn}

\usepackage{authblk}

\usepackage{mathptmx}       
\usepackage{helvet}         
\usepackage{courier}        
\usepackage{type1cm}        
\usepackage{makeidx}         
\usepackage{graphicx}        
\usepackage{multicol}        
\usepackage[bottom]{footmisc}

\usepackage{ucs}              

\usepackage{latexsym}
\usepackage{amsmath,amssymb,amscd,amsbsy}
\usepackage{upgreek}

\usepackage{tensor}

\usepackage{makeidx}         
\usepackage{graphicx}        
\usepackage{multicol}        
\usepackage[bottom]{footmisc}

\usepackage{rcs} \RCS $Revision: 2.2 $ \RCS $Date: 2018/09/02 11:16:02 $ \RCS $Author: hgm $ \RCS $RCSfile: 18_RV-algebra-model.tex,v $ \RCS $Id: 18_RV-algebra-model.tex,v 2.2 2018/09/02 11:16:02 hgm Exp $

\usepackage{anyfontsize}

\usepackage{mathrsfs}
\usepackage{bbm}
\usepackage{amsfonts}
\usepackage{amsmath}
\usepackage{enumerate}

\newcommand{\ignore}[1]{}

\newcommand{\vek}[1]{\mathchoice{\displaystyle\boldsymbol{#1}}
{\textstyle\boldsymbol{#1}}{\scriptstyle\boldsymbol{#1}}
{\scriptscriptstyle\boldsymbol{#1}}}

\newcommand{\tnb}[1]{\mathchoice{\displaystyle\mathboldsans{#1}}
{\textstyle\mathboldsans{#1}}{\scriptstyle\mathboldsans{#1}}
{\scriptscriptstyle\mathboldsans{#1}}}

\newcommand{\grad}{\mathop{\mathrm{grad}}\nolimits}

\usepackage[sfdefault=cmbr,OMLmathsans]{isomath}  
\usepackage{latexsym}
\usepackage{amsmath}
\usepackage{amsthm}
\usepackage{amssymb}
\usepackage{upgreek}  
\usepackage{tensor}  

\usepackage{float}
\usepackage{xcolor}

\makeindex             

\begin{document}


\title{Effect of Load Path on Parameter Identification for Plasticity Models using Bayesian Methods}
\author{{\small Ehsan Adeli$^{*}$, Bojana Rosi\'c$^{*}$, Hermann G. Matthies$^{*}$ and Sven Reinst\"{a}dler$^{\dag}$}}
\affil{{\small $^{*}$ Institute of Scientific Computing\\ $^{\dag}$ Institute of Structural Analysis\\ Technische Universit{\"a}t Braunschweig\\
Braunschweig, Germany\\ 
e.adeli@tu-braunschweig.de}}
\date{July 2018}

%
%


\maketitle

\abstract{To evaluate the cyclic behavior under different loading conditions using the kinematic and isotropic hardening theory of steel, a Chaboche viscoplastic material model is employed. The parameters of a constitutive model are usually identified by minimization of the distance between model response and experimental data. However, measurement errors and differences in the specimens lead to deviations in the determined parameters. In this article, the Choboche model is used and a stochastic simulation technique is applied to generate artificial data which exhibit the same stochastic behavior as experimental data. Then the model parameters are identified by applying an estimation using Bayes's theorem. The Gauss-Markov-Kalman filter using functional approximation is introduced and employed to estimate the model parameters in the Bayesian setting. Identified parameters are compared with the true parameters in the simulation, and the efficiency of the identification method is discussed. In the end, the effect of the load path on parameter identification is investigated.}

\section{Introduction}
\label{sec:Introduction}
In order to predict the behavior of loaded metallic materials, constitutive models are applied, which present a mathematical frame for the description of elastic and inelastic deformation. The models by Miller, Krempl, Korhonen,  Aubertin, Chan, and Bodner are well-known constitutive models for isotropic materials \cite{Miller, Krempl, Korhonen1, Aubertin, Chan}. In 1983, Chaboche \cite{Chaboche, Chaboche1} put forward what has become known as the unified Chaboche viscoplasticity constitutive model, which has been widely accepted.
	
All inelastic constitutive models contain parameters which have to be identified for a given material from experiments. In the literature only few investigations can be found dealing with identification problems using stochastic approaches. Klosowski and Mleczek have applied the least-squares method in the Marquardt-Levenberg variant to estimate the parameters of an inelastic model \cite{KLOSOWSKI}. Gong et al. have also used some modification of the least-squares method to identify the parameters \cite{Gong}. Harth and Lehn identified the model parameters of a model by employing some generated artificial data instead of experimental data using a stochastic technique \cite{Harth}. A similar study by Harth and Lehn has been done for other constitutive models like Lindholm and Chan \cite{Lindholm}.

In this paper, a viscoplastic model of Chaboche is studied. The model contains five material parameters which have to be determined from experimental data. It should be noted that here virtual data are employed instead of real experimental data. A cyclic tension-compression test is applied in order to extract the virtual data.

The model is described in Section 2, whereas Section 3 explains how to propagate the uncertainty in the model and how to perform the update. The probabilistic model is reformulated from the deterministic model, and once the forward model is provided, the model parameters are updated using a Bayesian approach.

In Section 4 the desired parameters are identified from the measured data. In fact, the parameters which have been considered as uncertain parameters are updated and the uncertainties of the them are reduced while the random variables representing the uncertain parameters are updated during the process. The results are thoroughly studied and the identified parameters as well as the corresponding model responses are analyzed. Finally the prediction of the models is compared with the measured data for different applied load paths. It is also explained why different load paths cause different identification of model parameters. 

\section{Model problem}
The mathematical description of metals under cyclic loading beyond the yield limit that includes viscoplastic material behavior as well as the characterization of compulsory isotropic-kinematic hardening is here given in terms of a modified Chaboche model introduced in \cite{dinkler}. As we consider classical infinitesimal strains, we assume an additive strain decomposition. The material behavior is described for the elastic part by isotropic homogeneous elasticity, and for viscoplasticity the dissipation potential is given by
\begin{equation}
     \phi(\vek{\sigma}) = \frac{k}{n+1}\Biggl\langle\frac{\sigma_{ex}}{k}\Biggr\rangle^{n+1},
     \label{eq:5}
\end{equation}
with $\langle\cdot\rangle = \max(0, \cdot)$ and $k$ and $n$ as the material parameters. Here $\sigma_{ex}$ is the over-stress, defined via the equivalent stress ($\sigma_{eq}$) which reads
\begin{equation}
\sigma_{eq}= \sqrt{\frac{3}{2} \text{tr} ((\vek{\sigma}-\vek{\chi})_D . ((\vek{\sigma}-\vek{\chi})_D)},
\end{equation}
where $(\cdot)_D$ denotes the deviatoric part and $\vek{\chi}$ is the back-stress of kinematic hardening. The over-stress $\sigma_{ex}$ is given by
\begin{equation}
    \sigma_{ex} = \sigma_{eq}-\sigma_y -R= \sqrt{\frac{3}{2} \text{tr} ((\vek{\sigma}-\vek{\chi})_D . (\vek{\sigma}-\vek{\chi})} - \sigma_y - R,
\end{equation}
where $\sigma_y$ is the yield stress and $R$ models the isotropic hardening which is introduced in the following. The partial derivative of the dissipation potential $\phi$ with respect to $\vek{\sigma}$ leads to the equation for the inelastic strain rate
\begin{equation}
    \dot{\vek{\epsilon}}_{vp} = \frac{\partial \phi}{\partial \vek{\sigma}} = \Biggl\langle\frac{\sigma_{ex}}{k}\Biggr\rangle^{n} \frac{\partial \sigma_{ex}}{\partial \vek{\sigma}}.
\end{equation}
The viscoplastic model allows for isotropic and kinematic hardening, which is considered in order to describe different specifications. Assuming $R(t)$ and $\vek{\chi}(t)$ with $R(0) = 0$ and $\vek{\chi}(0) = 0$ to describe isotropic and kinematic hardening respectively, the evaluation equations for these two are
\begin{equation}
\dot{R} = b_R(H_R-R)\dot{p}
\end{equation}
and
\begin{equation}
\dot{\vek{\chi}} = b_{\vek{\chi}}(\frac{2}{3} H_{\vek{\chi}}\frac{\partial \sigma_{eq}}{\partial \vek{\sigma}} - \vek{\chi} )\dot{p}
\end{equation}
respectively. In the evaluation equations of the both hardening, $\dot{p}$ is the viscoplastic multiplier rate given as:
\begin{equation}
    \dot{p} = \Biggl\langle\frac{\sigma_{ex}}{k}\Biggr\rangle^{n},
\end{equation}
which describes the rate of accumulated plastic strains. The parameter $b_R$ indicates the speed of stabilization, whereas the value of the parameter $H_R$ is an asymptotic value according to the evolution of the isotropic hardening. Similarly, the parameter $b_{\vek{\chi}}$ denotes the speed of saturation and the parameter $H_{\vek{\chi}}$ is the asymptotic value of the kinematic hardening variables. The complete model is stated in Table~\ref{tab:1}. Note that $\tnb{E}$ represents the elasticity tensor.

\begin{table}[H]
\caption{The constitutive model of Chaboche}
\begin{center}
\begin{minipage}{0.8\textwidth}
 \begin{tabular}{|| l||}   
 \hline
 \label{tab:1}
 Strain\\
 \centerline{$\vek{\epsilon}(t) = \vek{\epsilon}_e(t) + \vek{\epsilon}_{vp}(t)$} \\ \\
 Hooke's Law\\
 \centerline{$\vek{\sigma}(t) = \tnb{E}: \vek{\epsilon}_e(t)$}\\ \\
 Flow Rule\\
 \centerline{$\dot{\vek{\epsilon}}_{vp}(t) = 
     \langle\frac{\sigma_{eq}(t)-\sigma_y-R(t)}{k}\rangle^n \frac{\partial \sigma_{ex}}{\partial \vek{\sigma}}$}  \\ \\
 Hardening\\
  \centerline{$\dot{R} = b_R(H_R-R)\dot{p}$}\\ \\
 \centerline{$\dot{\vek{\chi}} = b_{\vek{\chi}}(\frac{2}{3} H_{\vek{\chi}}\frac{\partial \sigma_{eq}}{\partial \vek{\sigma}} - \vek{\chi} )\dot{p}$}\\ \\

 Initial Conditions\\
  \centerline{$\vek{\epsilon}_{vp}(0)=0$,~~$R(0)=0$,~~$\vek{\chi}(0)=0$}\\ \\
  
 Parameters\\
   \centerline{ $\sigma_y$ ~(Yield Stress)}\\ 
   \centerline{ $k$, $n$ ~(Flow Rule)}\\  
   \centerline{ $b_R$, $H_R$, $b_{\vek{\chi}}$, $H_{\vek{\chi}}$~(Hardening)}\\
 [1ex] 
 \hline
\end{tabular}
\end{minipage}
\end{center} 
\end{table}

By gathering all the desired material parameters to identify into the vector $\vek{q}=[\kappa~ G~ b_R~ b_{\vek{\chi}}~ \sigma_y]$, where $\kappa$ and $G$ are bulk and shear modulus, respectively, which determine the isotropic elasticity tensor, the goal is to estimate $\vek{q}$ given measurement displacement data, i.e. 
\begin{equation}
u=Y(\vek{q})+e
\label{eq:pce_uf_yf},
\end{equation}
in which $Y(\vek{q})$ represents the measurement operator and $e$ the measurement (also possibly the model) error. Being an ill-posed problem, the estimation of $\vek{q}$ given $u$ is not an easy task and usually requires regularization. This can be achieved either in a deterministic or a probabilistic setting. Here, the latter one is taken into consideration as further described in the text.

\section{Bayesian identification}
By using additional (prior) knowledge on the parameter set next to the observation data, the probabilistic approach 
regularizes the problem of estimating $\vek{q}$ with the help of Bayes's theorem
\begin{equation}
\label{eq:bayesrule}
\pi_{q|u}(\vek{q}|u)\propto L(u|\vek{q}) \pi_q(\vek{q}),
\end{equation}
in which the likelihood $L(u|\vek{q})$ describes how likely the measurement data are given prior knowledge $\pi_q(\vek{q})$. This in turn requires the reformulation of the deterministic model into a probabilistic one, and hence the propagation of material uncertainties through the model ---the so-called forward problem--- in order to obtain the likelihood \cite{Matthies0,Matthies01}. 

The main difficulty in using Equation (\ref{eq:bayesrule}) lies in the computation of the likelihood. Various numerical algorithms can be applied, the most popular example of which are the Markov chain Monte Carlo methods. Being constructed on the fundamentals of ergodic Markov theory, these methods are characterized by very slow convergence. To avoid this, an approximate method based on Kolmogorov's definition of conditional expectation as already presented in \cite{Matthies2} is considered here. 

Let the material parameters $\vek{q}$ be modeled as random variables on a probability space  $S:=L_2(\Omega,\mathcal{B},\mathbb{P})$. Here, $\Omega$
denotes the space of elementary events $\omega$, $\mathcal{B}$ is the $\sigma$-algebra and $\mathbb{P}$ stands for the probability measure. This alternative formulation of Bayes's rule can be achieved by expressing the conditional probabilities in Equation (\ref{eq:bayesrule}) in terms of conditional expectation. Following the mathematical derivation in \cite{Matthies2,Matthies3,Bojana,Bojana1}, this approach boils down to a quadratic minimization problem by considering the forecast random variable $q_f$ and the update of the forecast random variable $q_a$:
\begin{equation}
\label{eq:minprob}
 \bar{q}^{|z}=P_{\mathscr{Q}_{sn}} q_f=\underset{\eta\, \in \, \mathcal{Q}_{sn}}{\textrm{arg min }} \|q_f-\eta\|^2_{L_2} = \vek{\Xi} (u_f(\omega)),
\end{equation}
where $P_{\mathcal{Q}_{sn}}$ is the orthogonal projection operator of $q_f$ onto the space of the new information $\mathcal{Q}_{sn}:=\mathscr{Q} \otimes S_n$ in which the space $S_n$ is the space of random variables generated by the measurement $u = Y(\vek{q}) + e$.
Due to the Doob-Dynkin lemma, $\bar{q}^{|z}$ is a function of the observation, where $u_f(\omega) = Y(q(\omega)) + e(\omega)$ is the forecast, and the assimilated value is $q_a(\omega) = q_f(\omega) + ( \vek{\Xi}(\hat{z}) - \vek{\Xi}(u_f(\omega)) )$. 

Constraining the space of all functions $\vek{\Xi}$ to the subspace of linear maps, the minimization problem in Equation (\ref{eq:minprob}) leads to a unique solution $K$. Note that the projection is performed over a smaller space than $\mathcal{Q}_{sn}$. An implication of this is that available information is not completely used  in the process of updating, introducing an approximation error.  This gives an affine approximation of Equation (\ref{eq:minprob})
\begin{equation}
\label{eq:gmarkovthm}
 q_a(\omega) = q_f(\omega) +  K(\hat{z}-u_f(\omega)),
\end{equation}
also known as a linear Bayesian posterior estimate or the so-called Gauss-Markov-Kalman filter (GMKF). Here, $q_f$ represents the prior random variable, $q_a$ is the posterior approximation, $u_f = Y(q_f(\omega)) + e(\omega)$ is the predicted measurement and $K$ represents the very well-known Kalman gain
\begin{equation}
 K:= C_{q_f u_f} \bigl(C_{u_f}+C_{\varepsilon}\bigr)^{-1},
 \label{eq:kalman1}
\end{equation}
which can be easily evaluated if the appropriate covariance matrices $C_{q_f u_f}$, $C_{u_f}$ and $C_{\varepsilon}$ are known. 

An advantage of Equation (\ref{eq:gmarkovthm}) compared to Equation (\ref{eq:bayesrule}) is that the inference in Equation~(\ref{eq:gmarkovthm}) is given in terms of random variables instead of conditional densities.
Namely, $q_a(\omega)$, $q_f(\omega)$, and $u_f(\omega)$ denote the random variables used to model the posterior, prior, observation, and predicted observation, respectively. 

In this light the linear Bayesian procedure can be reduced to a simple algebraic method. Starting from the functional representation of the prior
 \begin{equation}
  \hat{q}_f (\omega) = \sum_\alpha q_f^{(\alpha)}
  \psi_\alpha(\omega),
  \label{eq:q_f}
 \end{equation}
where $\psi_\alpha$ are multivariate Hermite polynomials, and by considering the proxy in Equation (\ref{eq:q_f}), one may discretize Equation (\ref{eq:gmarkovthm}) as:

\begin{equation}
\label{eq:lbunum}
 \vek{q}_a = \vek{q}_f +K\bigl(\hat{\vek{z}} - \vek{u}_f \bigr),
\end{equation}
where $\vek{q}_a = [..., q^{(n)}_a,...]$, etc. are the PCE coefficients. As the measurement is a deterministic value, $\hat{z}= [\hat{z},0,...,0]$ has only a zero-th order tensor. 
The covariances for the Kalman gain Equation \ref{eq:kalman1} are easily computed, e.g.
\begin{equation}
  C_{\hat{q}_f,\hat{u}_f} =\sum_{\alpha>0} \alpha ! \; q_f^{(\alpha)} (\vek{u}_f^{(\alpha)})^T.
\end{equation}

\section{Numerical results}

The identification of the material constants in the Chaboche unified viscoplasticity model is a reverse process, here based on virtual data. In case of the Chaboche model the best way of parameter identification is using the results of the cyclic tests, since more information can be obtained from cyclic test rather than creep and relaxation tests, specifically information regarding hardening parameters. The aim of the parameter identification is to find a parameter vector $\vek{q}$ introduced in the previous section. The bulk modulus ($\kappa$), the shear modulus ($G$), the isotropic  hardening coefficient ($b_{R}$), the kinematic hardening coefficient ($b_{\vek{\chi}}$) and the yield stress ($\sigma_y$) are considered as the uncertain parameters of the constitutive model. 

\begin{figure}[H]
    \begin{center}{}
     \includegraphics[width=2.55in]{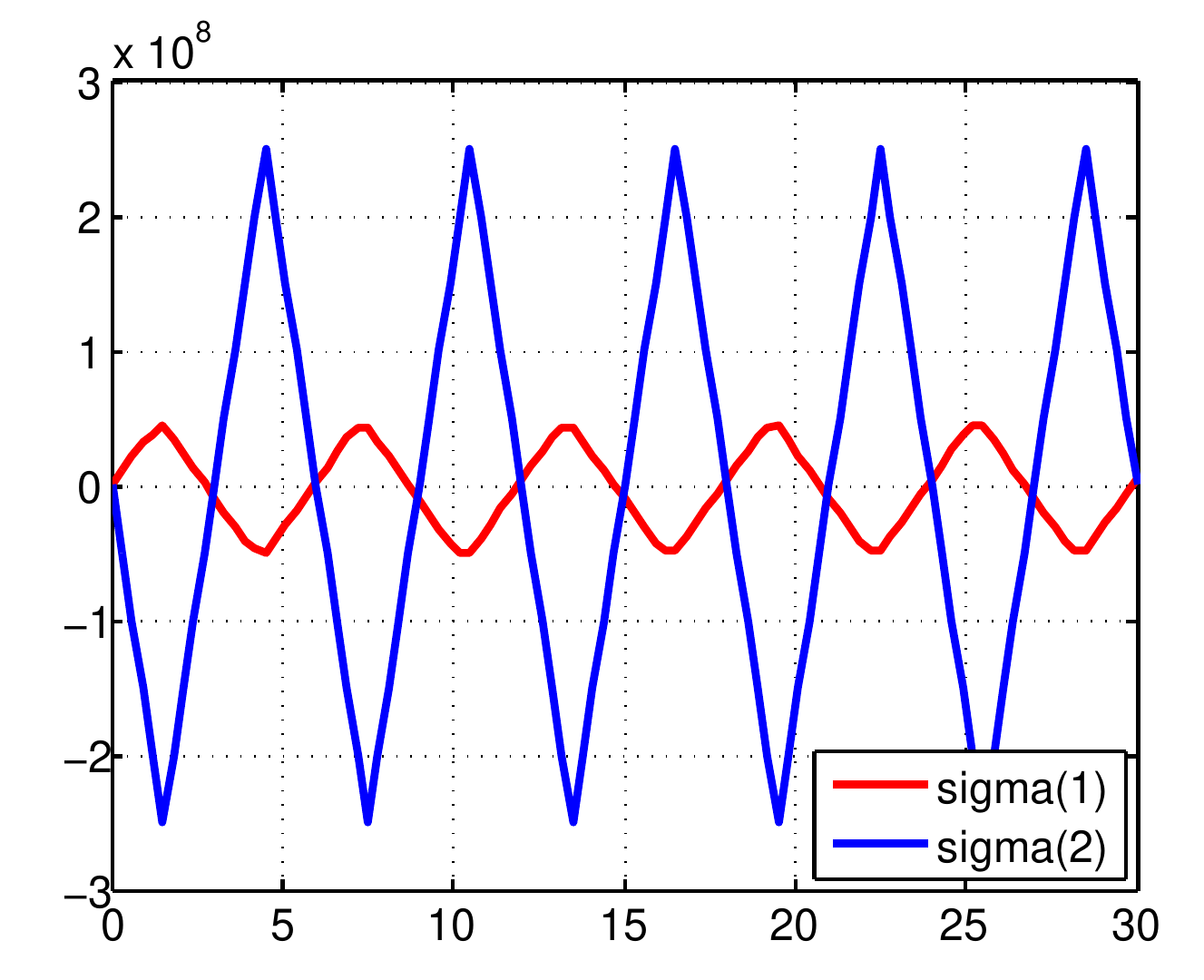}
     
     \caption{\label{fig:sigma-time} Decomposed applied force on desired node according to time- Case 1}
    \end{center}\hspace{2pc}%
\end{figure}

\begin{figure}[H]
    \begin{center}{}
     \includegraphics[width=2.55in]{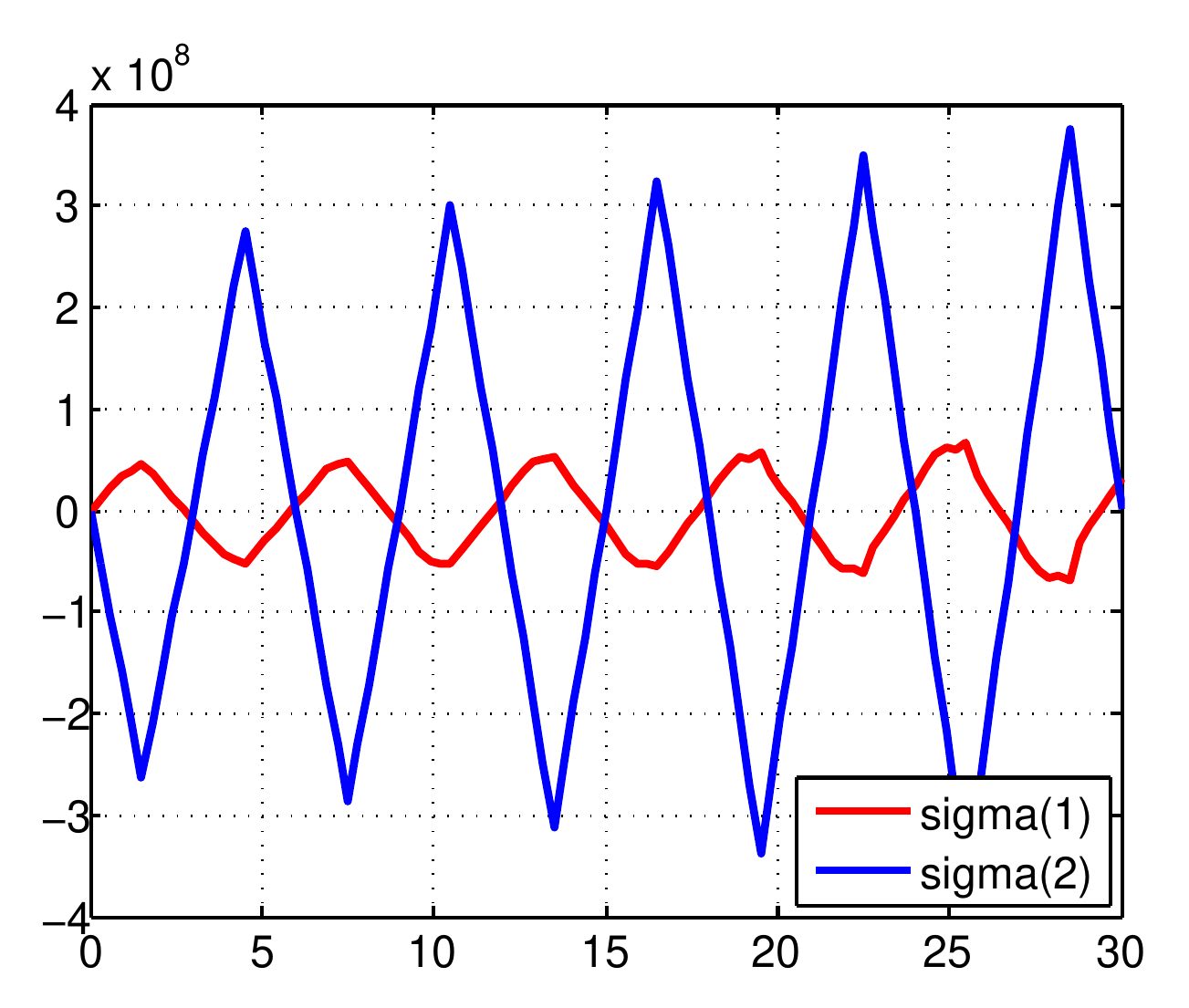}
     
     \caption{\label{fig:sigma-time1} Decomposed applied force on desired node according to time- Case 2}
    \end{center}\hspace{2pc}%
\end{figure}

\begin{figure}[H]
\centering
\includegraphics[width = 2.2in]{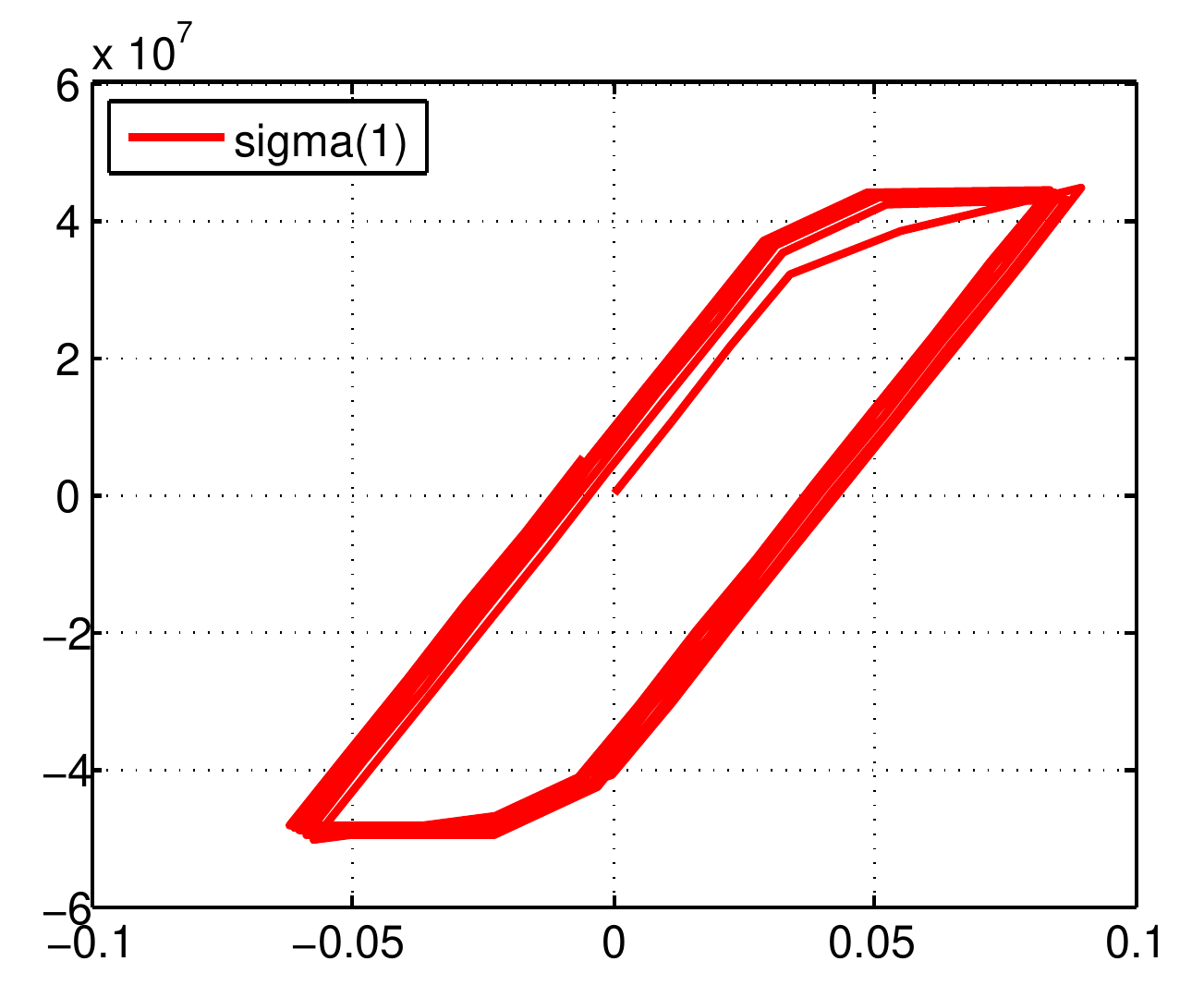}
$\quad$
\includegraphics[width = 2.2in]{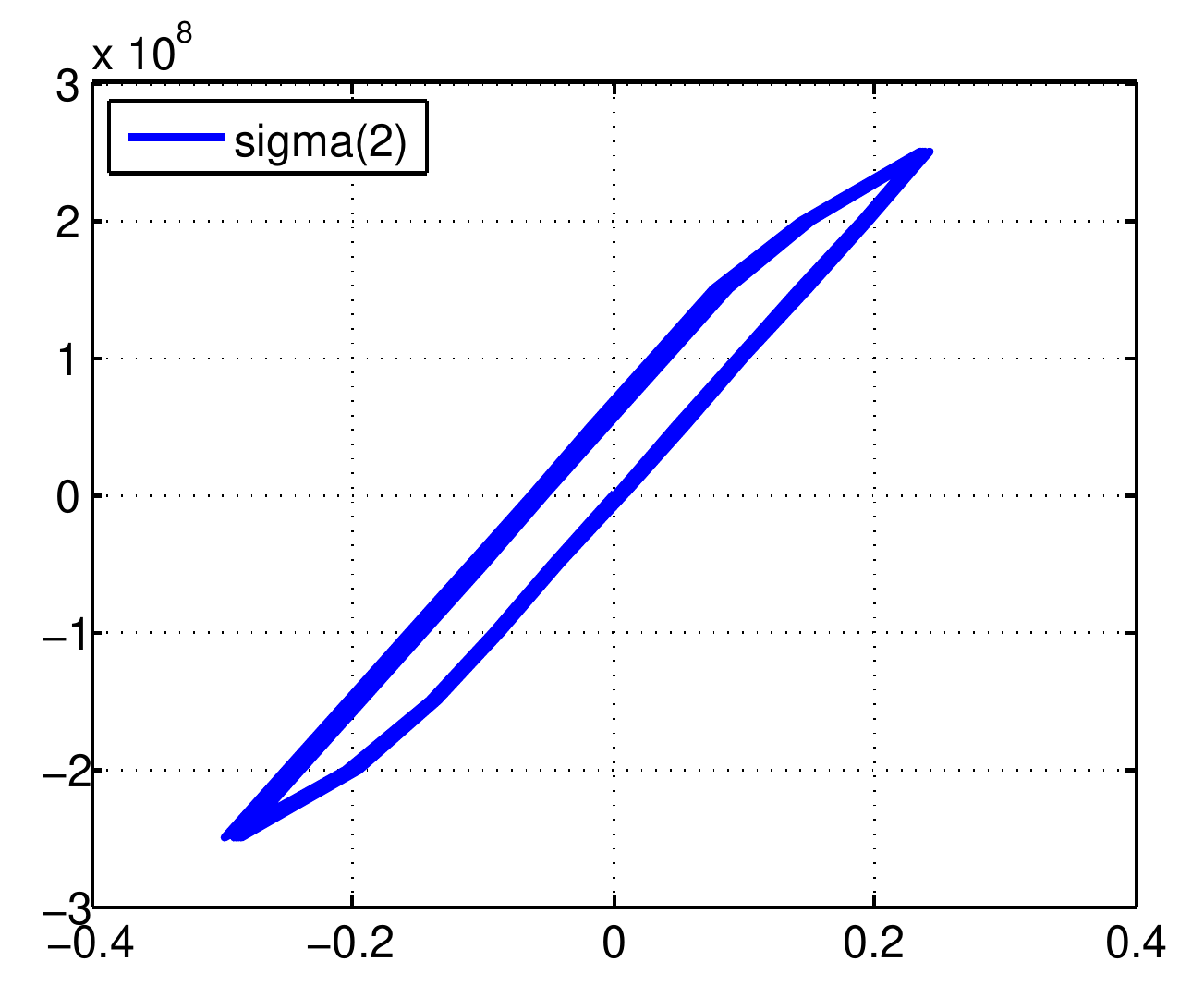}
\caption{\label{fig:sigma-epsilon} $\sigma$-$\epsilon$ for node on the front surface in plane and normal directions- Case 1}
\end{figure}

\begin{figure}[H]
\centering
\includegraphics[width = 2.2in]{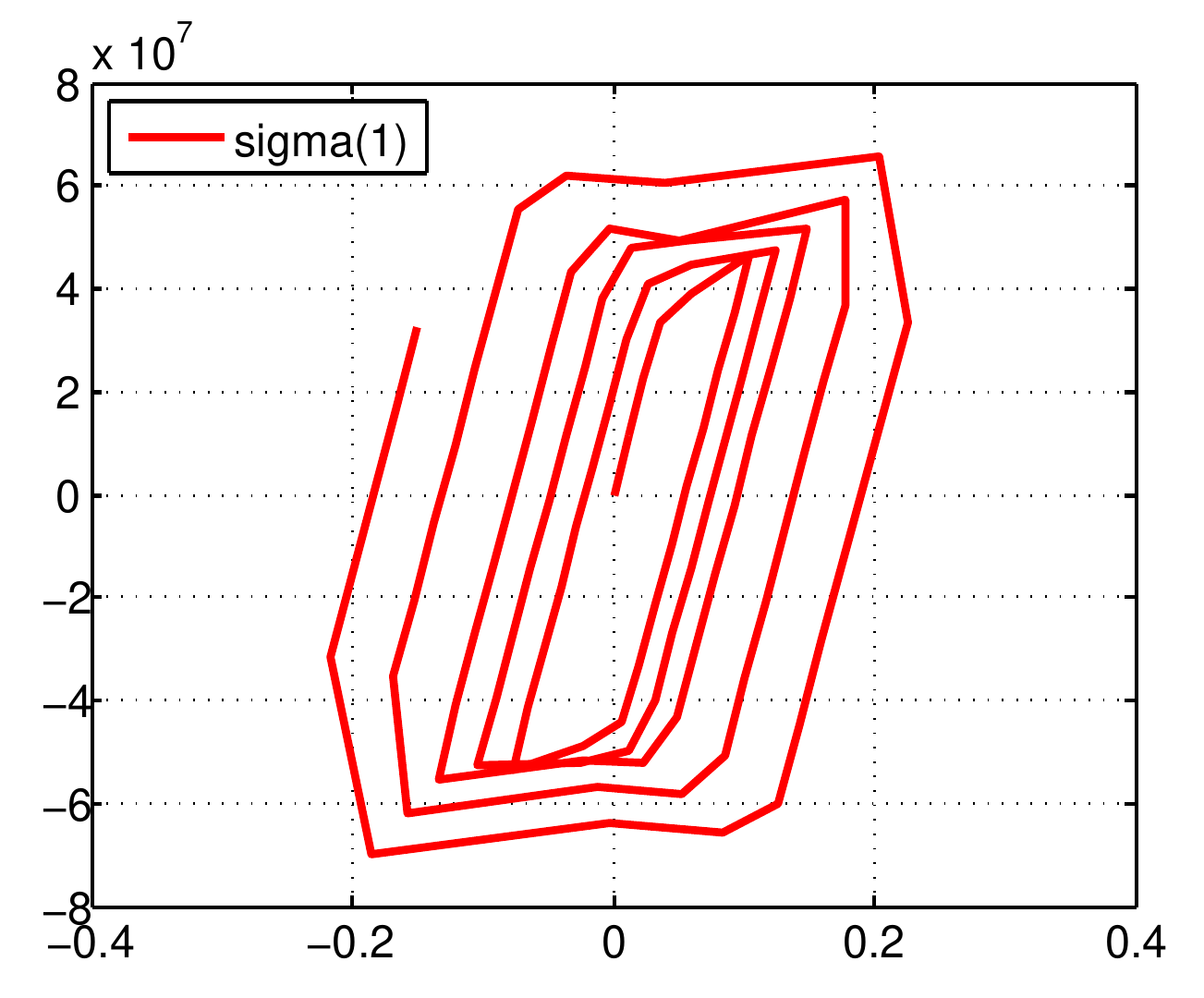}
$\quad$
\includegraphics[width = 2.2in]{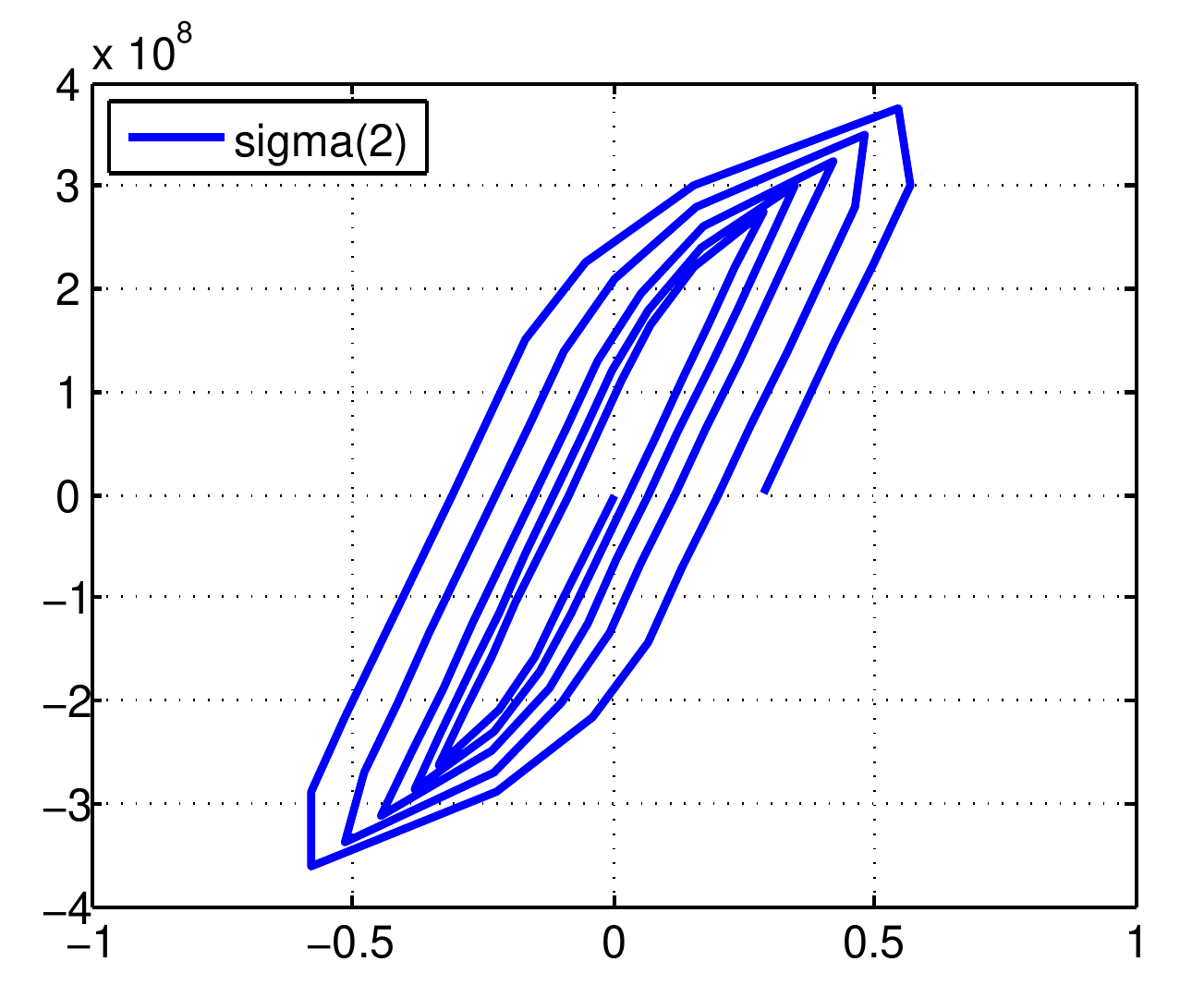}
\caption{\label{fig:sigma-epsilon1} $\sigma$-$\epsilon$ for node on the front surface in plane and normal directions- Case 2}
\end{figure}

A preliminary study is on a regular cube, modeled with one 8 node element, completely restrained on the back face, and with normal traction on the opposite (front) face. Two cases are considered in order to compare the effect of applied force on identified parameters. For both cases the magnitude of the normal traction and a stress in the plane of the front face are plotted in Fig.~\ref{fig:sigma-time} and Fig.~\ref{fig:sigma-time1}, respectively. Blue and red colors represent the stress value in normal and in plane directions, respectively. As it is seen, the magnitude of the applied force for the case 1 is constant all time but for the case 2 the magnitude of the applied force grows gradually by time.
\\
\vspace{-1.25cm}
Considering the parameters listed in Table~\ref{tab:general11}, the related $\sigma$-$\epsilon$ hysteretic graph obtained for the applied force case 1 and 2 which can be seen in Fig.~\ref{fig:sigma-epsilon} and Fig.~\ref{fig:sigma-epsilon1}, respectively. 
\newline
\newline

\begin{figure}[H]
\centering
\includegraphics[width = 2.2in]{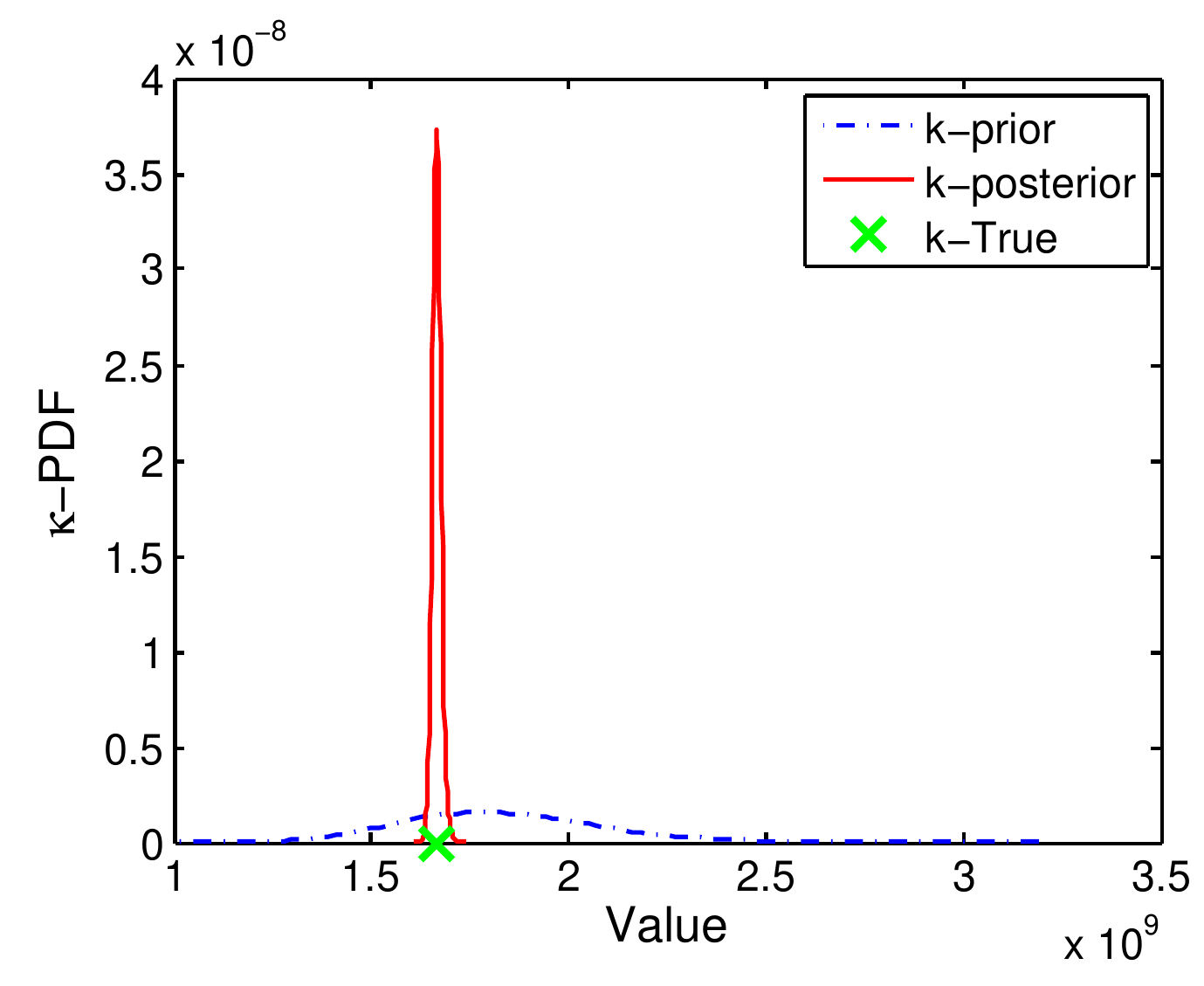}
$\quad$
\includegraphics[width = 2.2in]{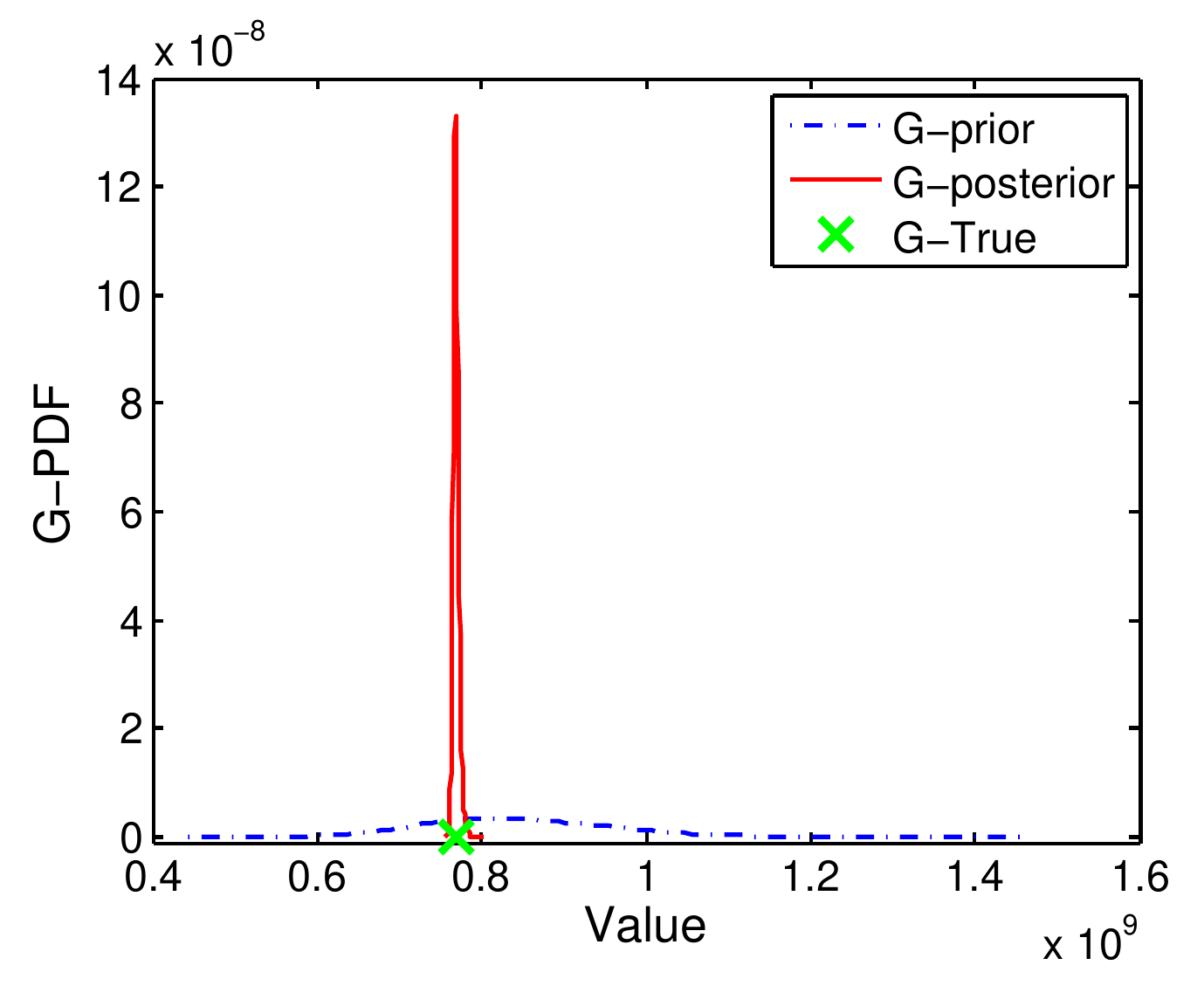}
$\quad$
\includegraphics[width = 2.2in]{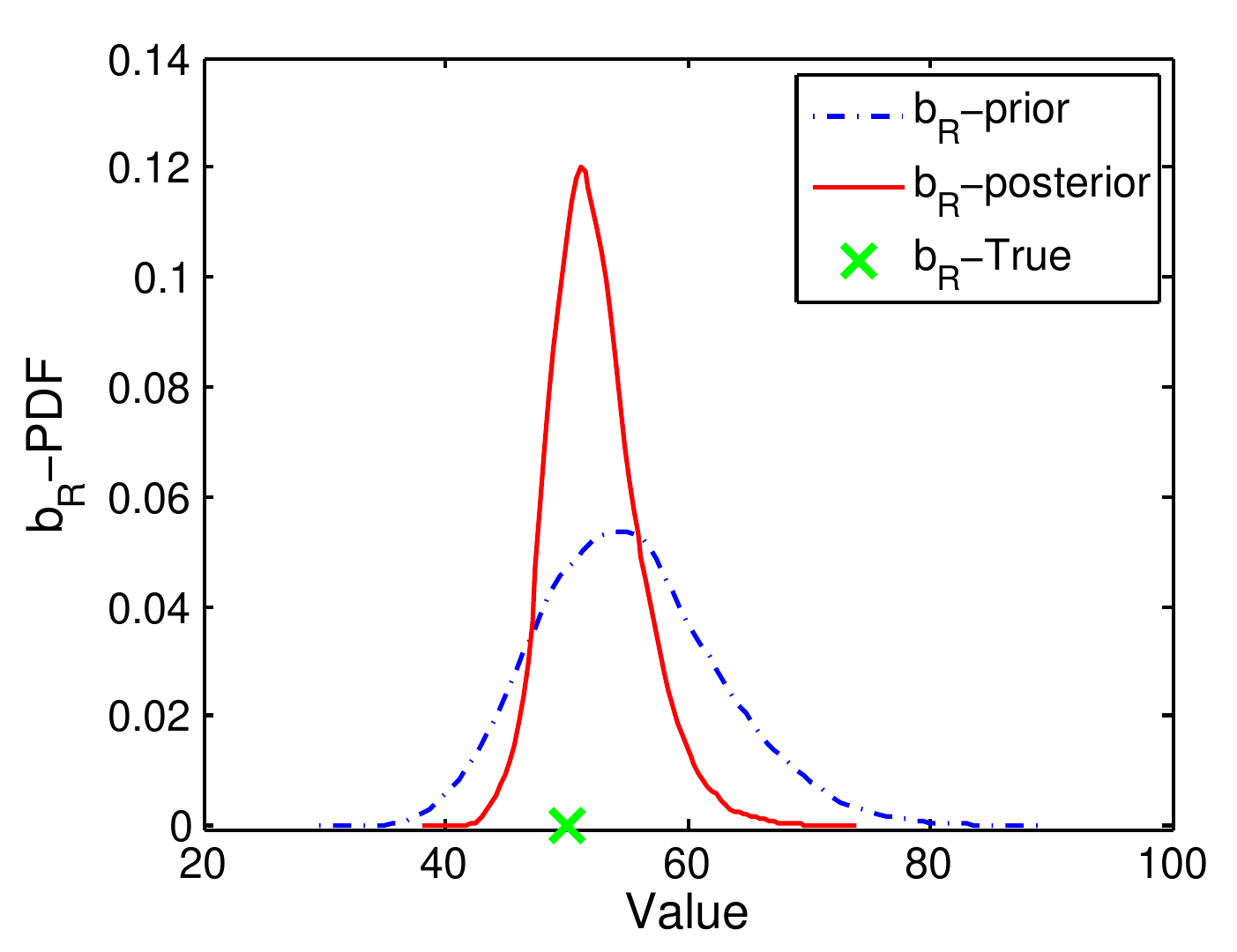}
$\quad$
\includegraphics[width = 2.2in]{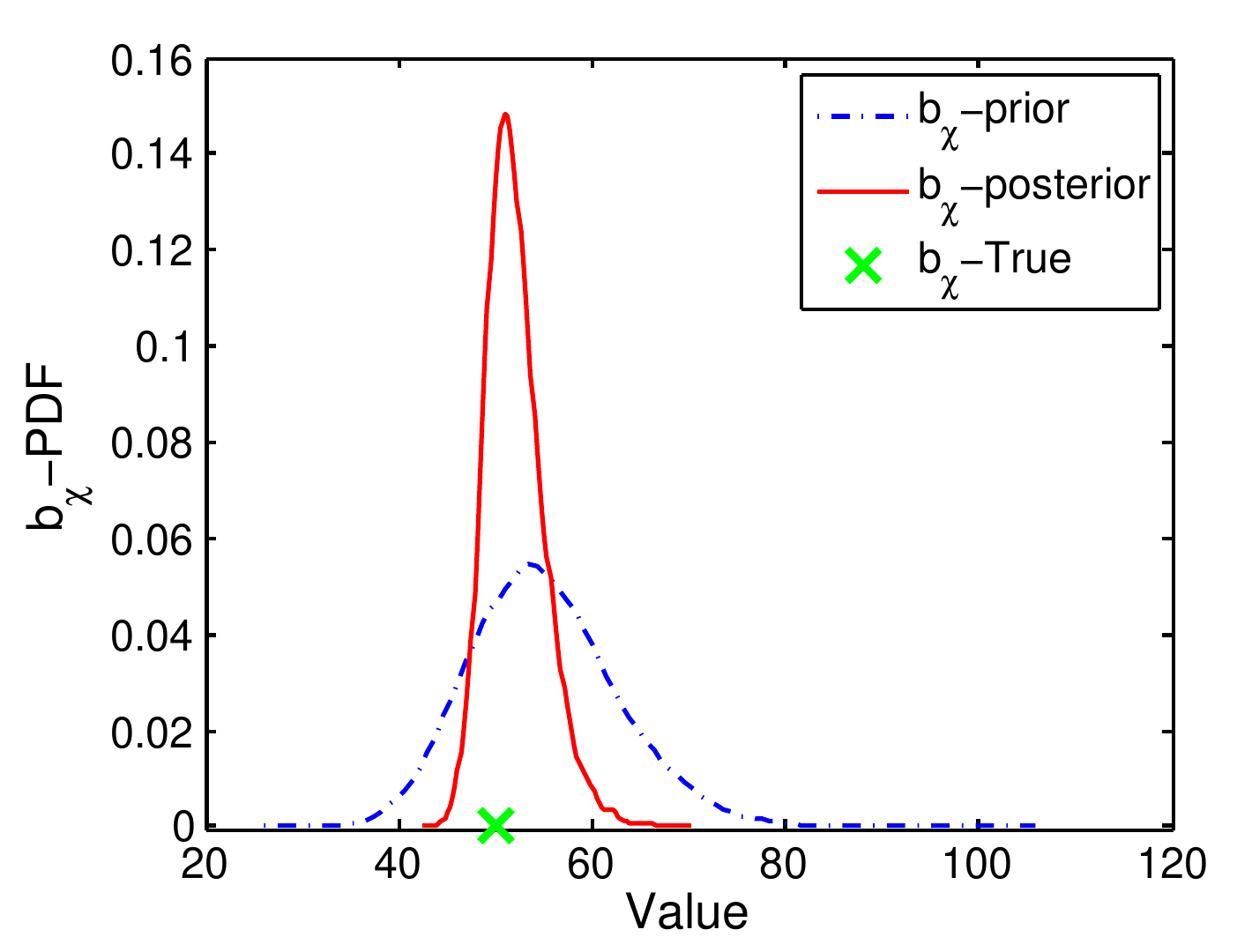}
$\quad$
\includegraphics[width = 2.2in]{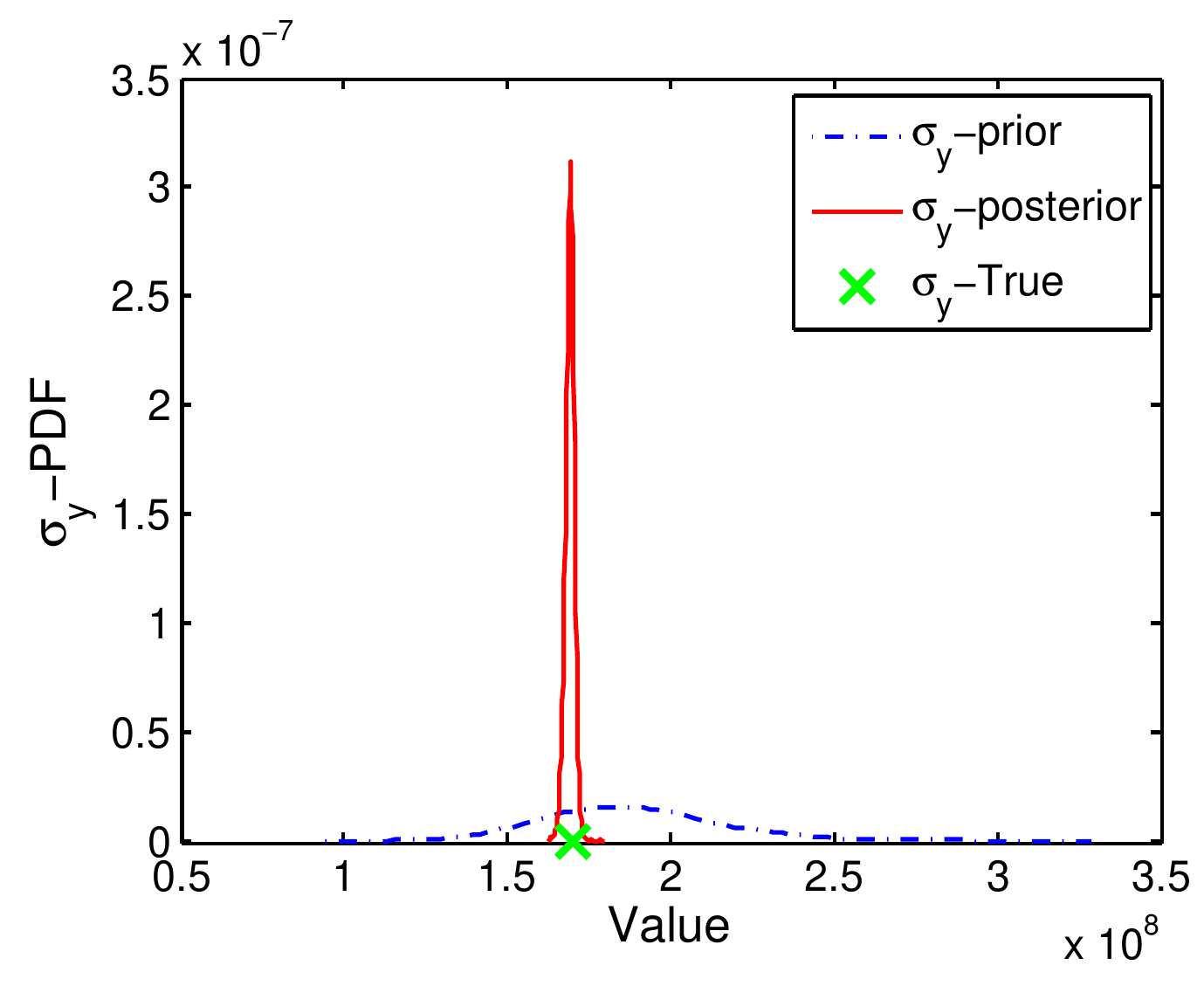}
\caption{\label{fig:parameters} PDF of identified parameters- Case 1}
\end{figure}

\begin{figure}[H]
\centering
\includegraphics[width = 2.2in]{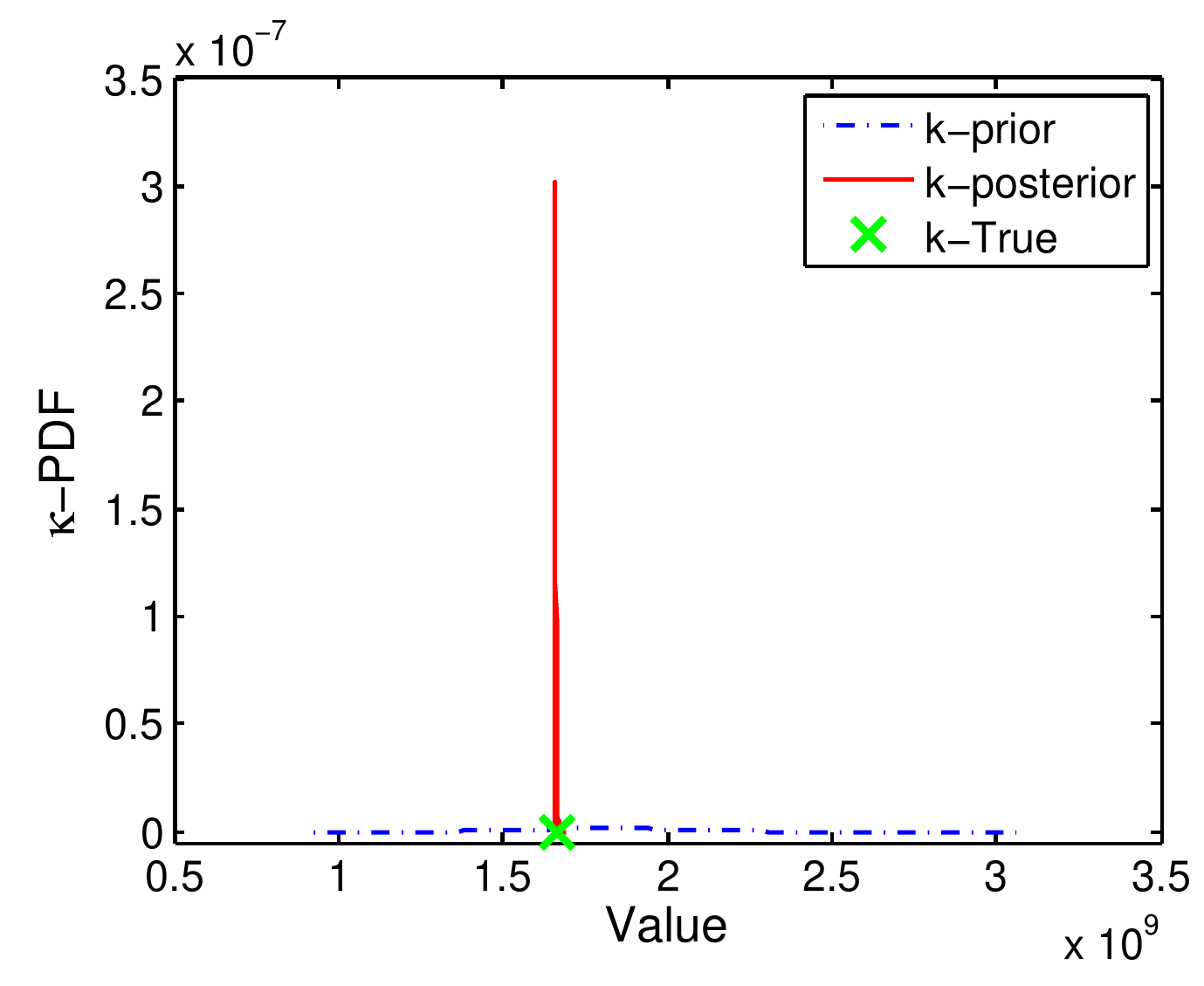}
$\quad$
\includegraphics[width = 2.2in]{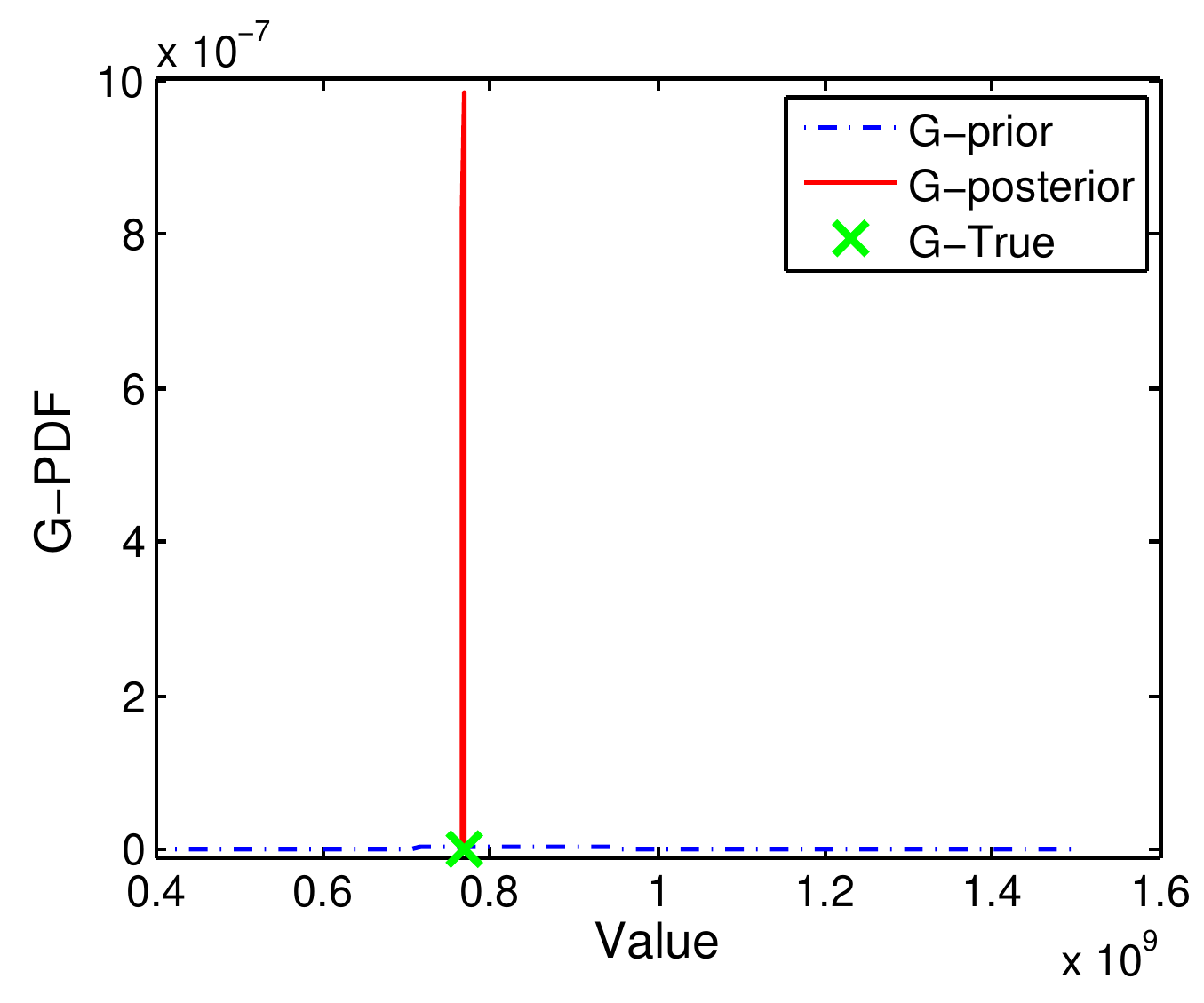}
$\quad$
\includegraphics[width = 2.2in]{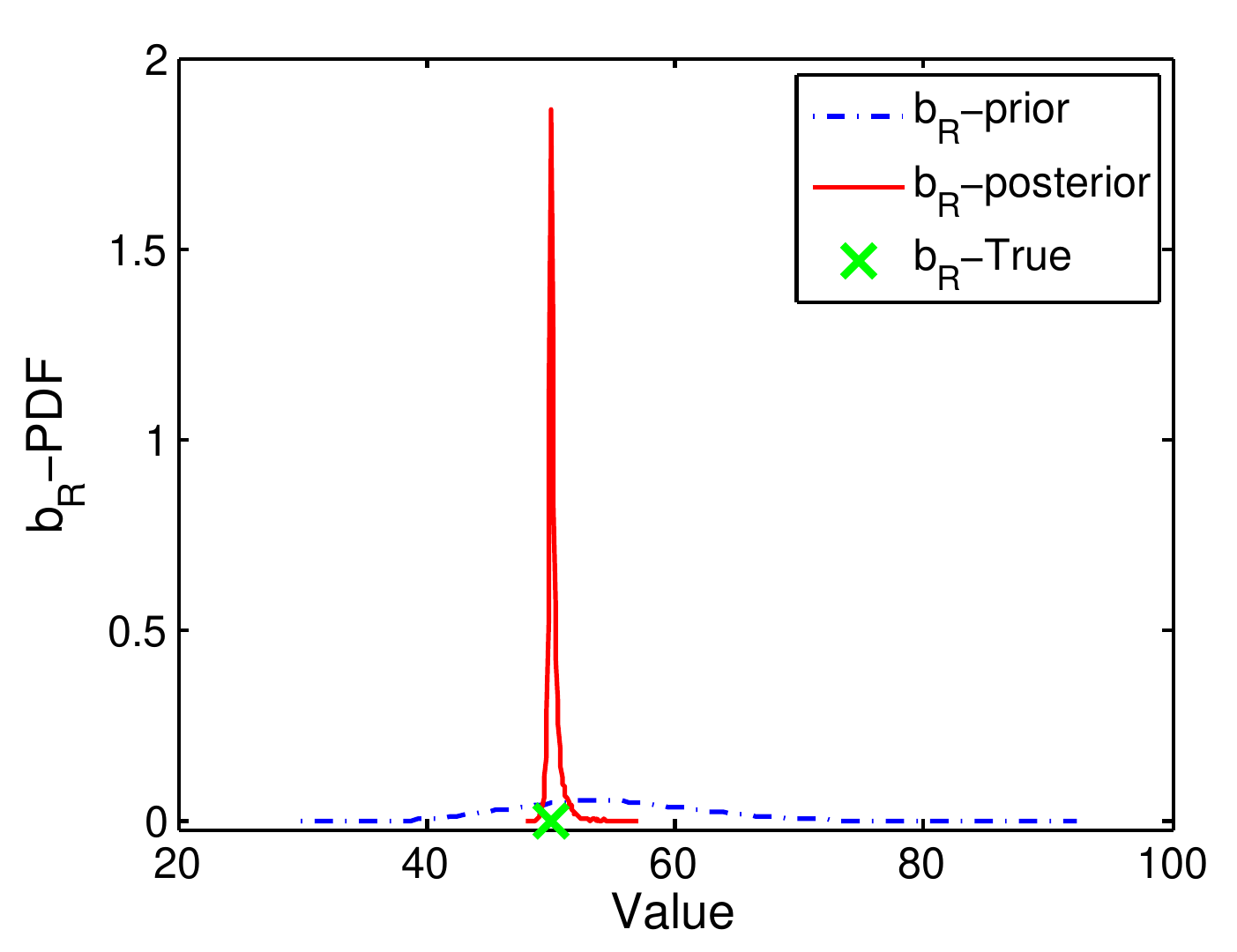}
$\quad$
\includegraphics[width = 2.2in]{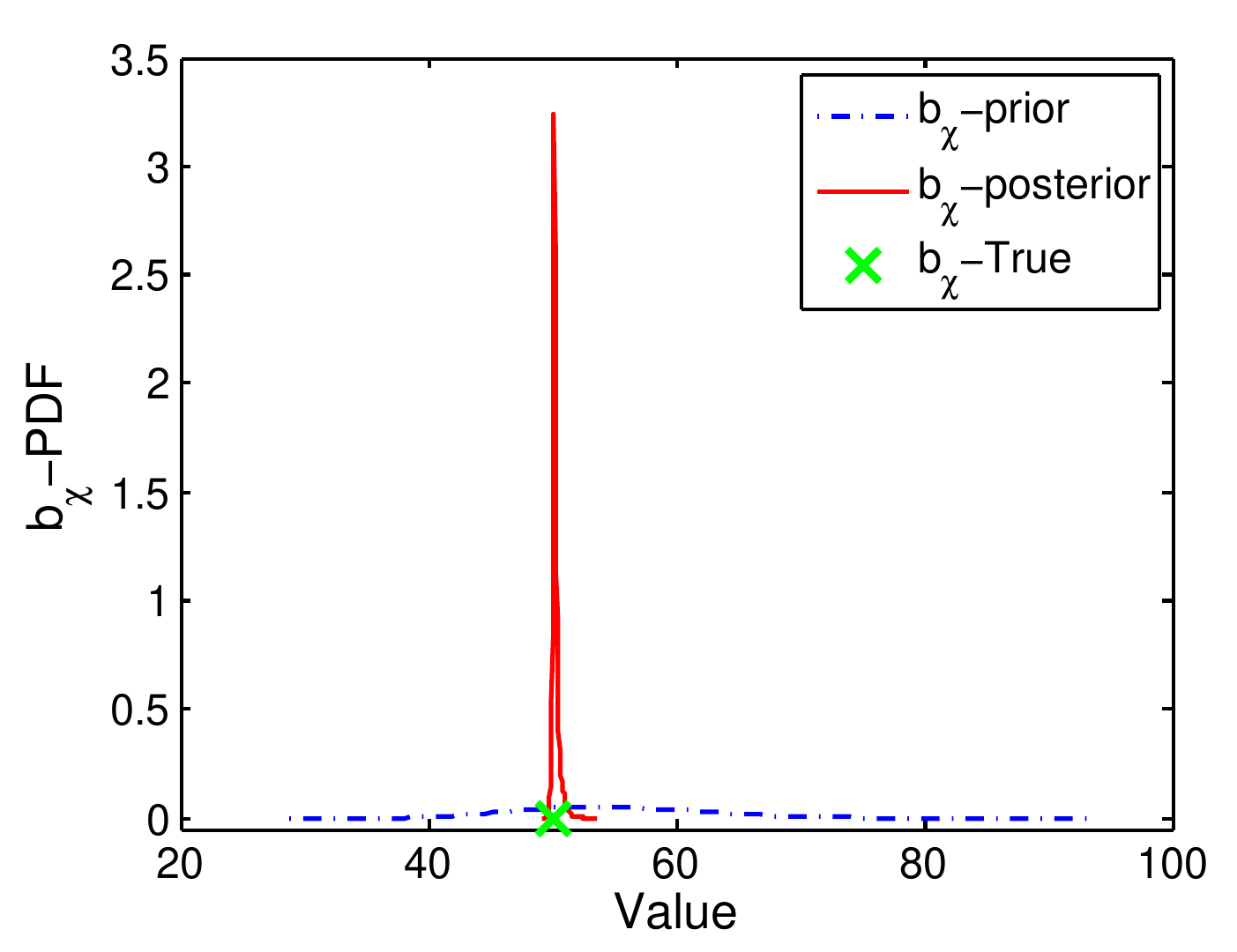}
$\quad$
\includegraphics[width = 2.2in]{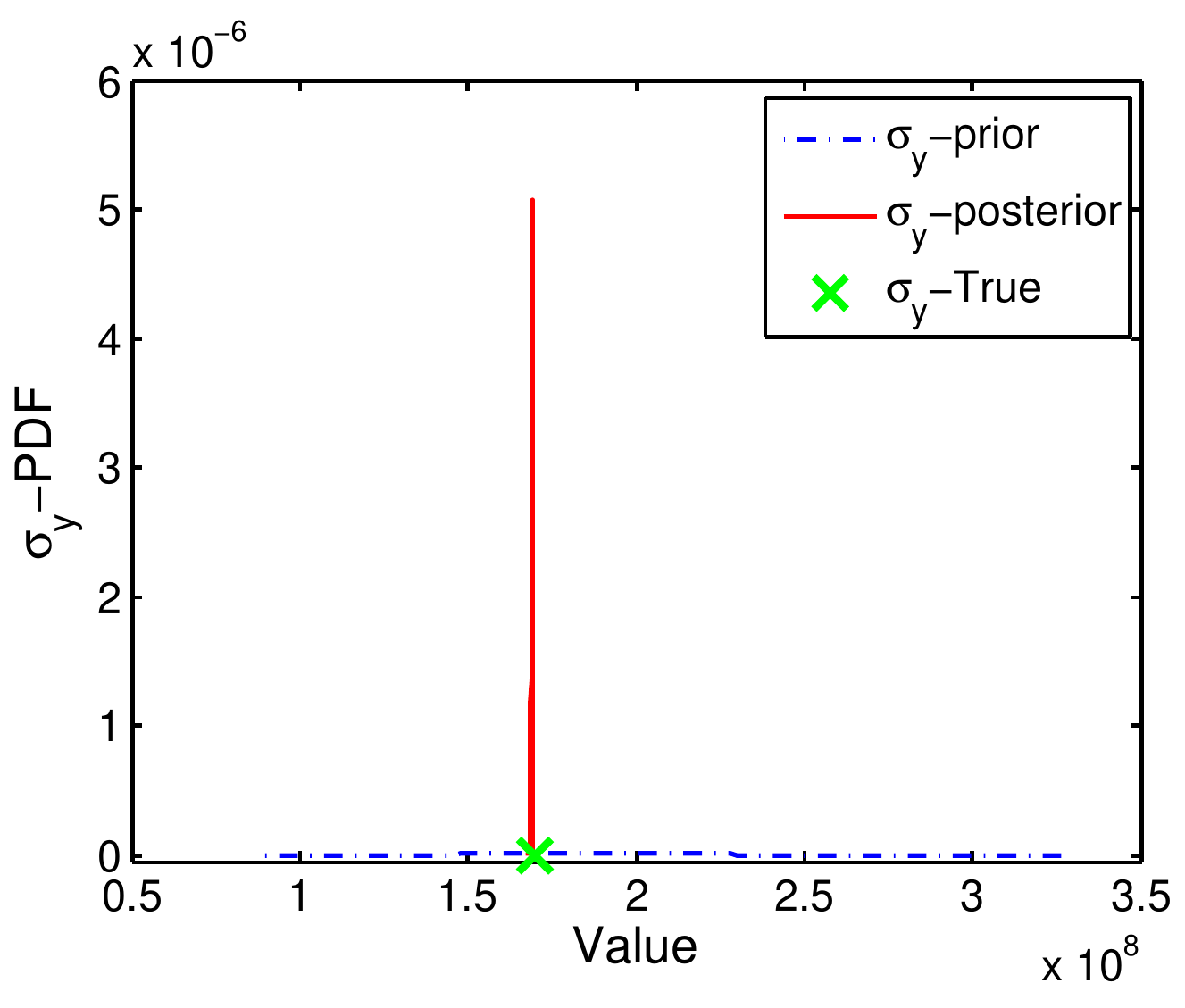}
\caption{\label{fig:parameters1} PDF of identified parameters- case 2}
\end{figure}

\begin{table}[H]
\caption{The model parameters}
\begin{center}
\begin{tabular}{l*{9}{c}r}
\hline
              & $\kappa$ & $G$ & $\sigma_y$ & $n$ & $k$ & $b_{R}$ & $H_{R}$ & $b_{\vek{\chi}}$ &  $H_{\vek{\chi}}$ &  \\
\hline
             & $1.66\mathrm{e}9$ & $7.69\mathrm{e}8$  &$1.7\mathrm{e}8$ & $1.0$ & $1.5\mathrm{e}8$ & $50$ & $0.5\mathrm{e}8$ & $50$ &  $0.5\mathrm{e}8$ &  \\
\hline
\label{tab:general11}

\end{tabular}
\end{center}
\end{table}
\vspace{-1.0cm}
The displacements of one of the nodes on the front surface in normal and in plane directions are observed as the virtual data in this study. Applying the Gauss-Markov-Kalman filter with functional approximation as explained in the previous chapter and introducing measurement error in such a way that 15 percent of mean values are equal to the coefficient of variation for the related parameter, the probability density function (PDF) of prior and posterior of the identified parameters can be seen in Fig.~\ref{fig:parameters} and Fig.~\ref{fig:parameters1} for the first and second case, respectively. 

Summarizing the results, the true values $q_{\text{true}}$ and the mean and standard deviation of the estimated parameters, $\vek{q}^m_{\text{est}}$ and $\vek{q}^{\text{std}}_{\text{est}}$ respectively, for both cases are compared in Table~\ref{tab:general41}.

\begin{table}[H]
\caption{The identified model parameters}
\begin{center}
\begin{tabular}{l*{6}{c}r }
\hline
Parameters      & $\vek{q}_{\text{true}}$  & $\vek{q}^m_{\text{est-\text{1}}}$&$\vek{q}^{\text{std}}_{\text{est-\text{1}}}$ & $\vek{q}^m_{\text{est-\text{2}}}$&$\vek{q}^{\text{std}}_{\text{est-\text{2}}}$ &\\
\hline
  $\kappa$ & $1.66\mathrm{e}9$ & $1.66\mathrm{e}9$ & $1.13\mathrm{e}7$ &  $1.66\mathrm{e}9$ & $2.59\mathrm{e}6$ &\\
  $G$             & $7.69\mathrm{e}8$  &$7.68\mathrm{e}8$ & $3.47\mathrm{e}6$ & $7.68\mathrm{e}8$ & $6.39\mathrm{e}5$ &\\
  $b_{R}$              & 50  & 52.36 & 3.71 & 50.27 & 0.29 & \\
  $b_{\vek{\chi}}$             & 50  & 52.04 & 3.01 & 50.19 & 0.53  &\\
  $\sigma_y$                  & $1.7\mathrm{e}8$  & $1.69\mathrm{e}8$ & $1.35\mathrm{e}6$ & $1.69\mathrm{e}8$ & $1.52\mathrm{e}5$ & \\
\hline
\end{tabular}\label{tab:general41}
\end{center}
\end{table}
\subsection{Discussion of the results}

From the sharpness of the posterior PDF of $\kappa$, $G$ and $\sigma_{y}$, it can be concluded that enough information from virtual data is received and updating the parameters considering their uncertainty is done very properly for the both cases, as the standard deviation of the residual uncertainty is below $1\%$ of the mean.

For the posterior PDF of $b_{R}$ and $b_{\vek{\chi}}$, it can be inferred that better updating is done for the second case compared to the first case. Not only are the more accurate estimations of the exact hardening parameters, $b_{R}$ and $b_{\vek{\chi}}$, predicted for the second case, but the uncertainty of the estimated hardening parameters are also reduced much more for the second case. 

\begin{figure}[H]
\centering
\includegraphics[width = 1.80in]{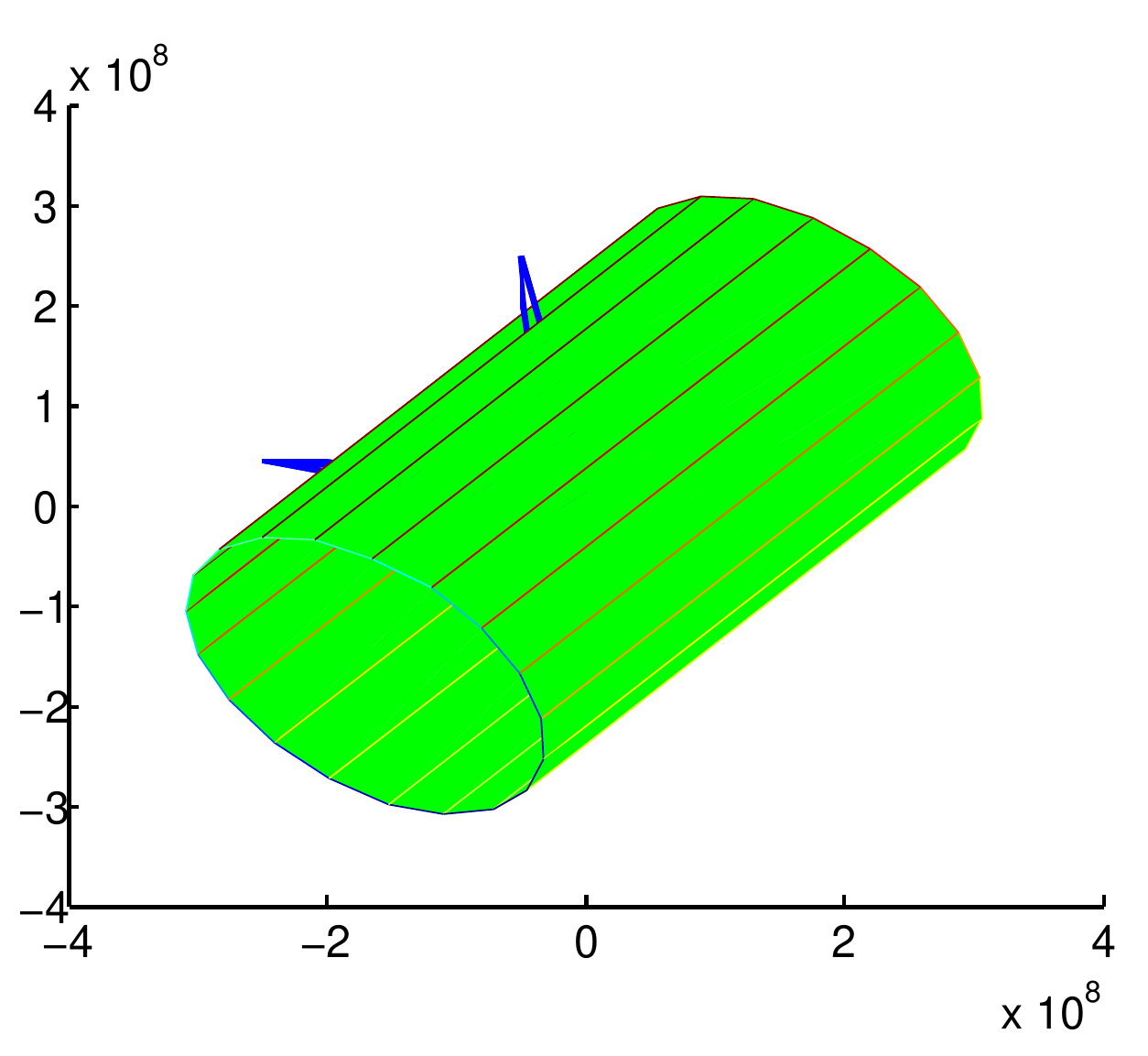}
$\quad$
\includegraphics[width = 2.60in]{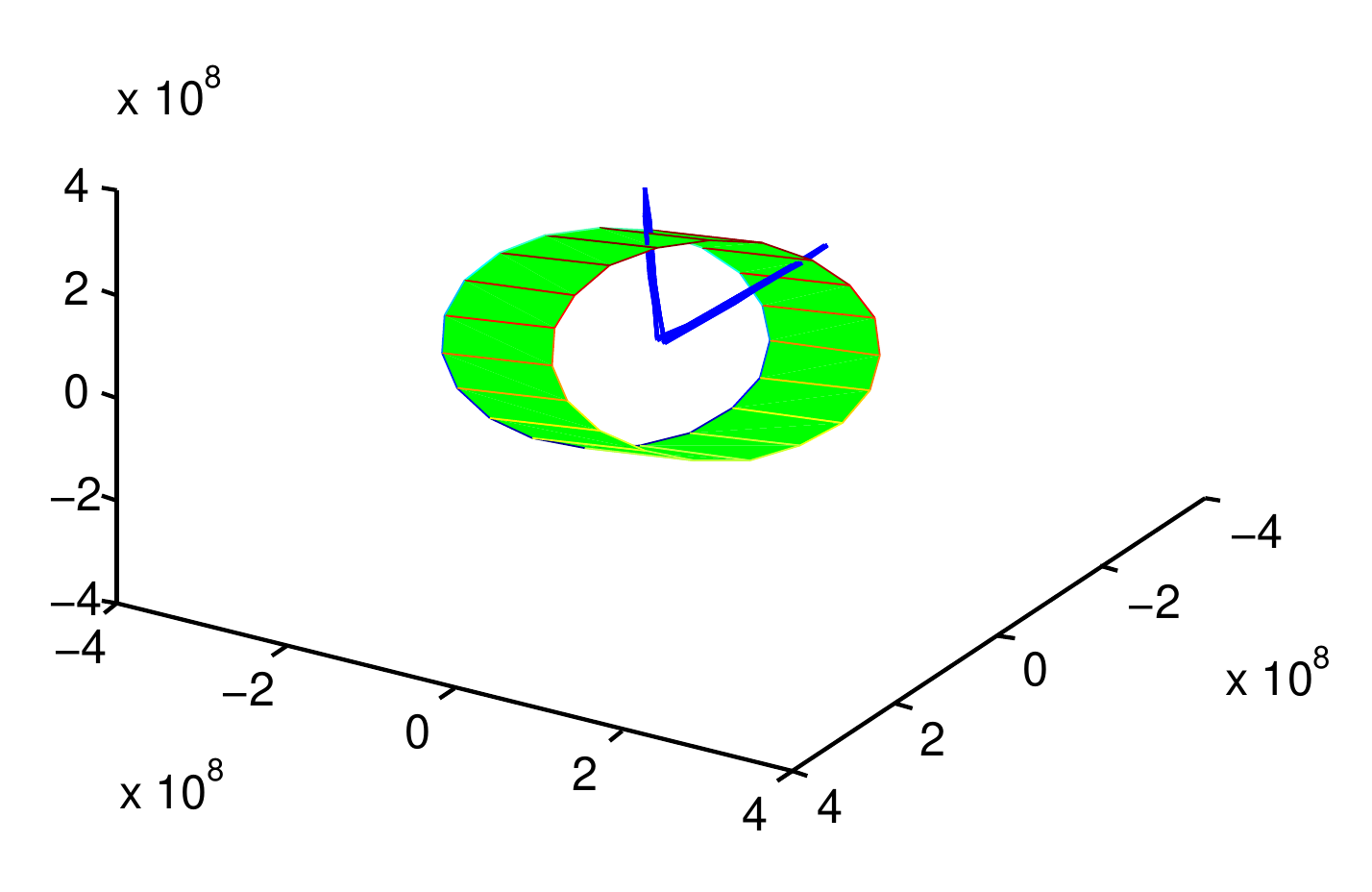}
\caption{\label{fig:VM} Principal stresses of applied force in 3D considering the von Mises yield criterion- Case 1}
\end{figure}

\begin{figure}[H]
\centering
\includegraphics[width = 1.80in]{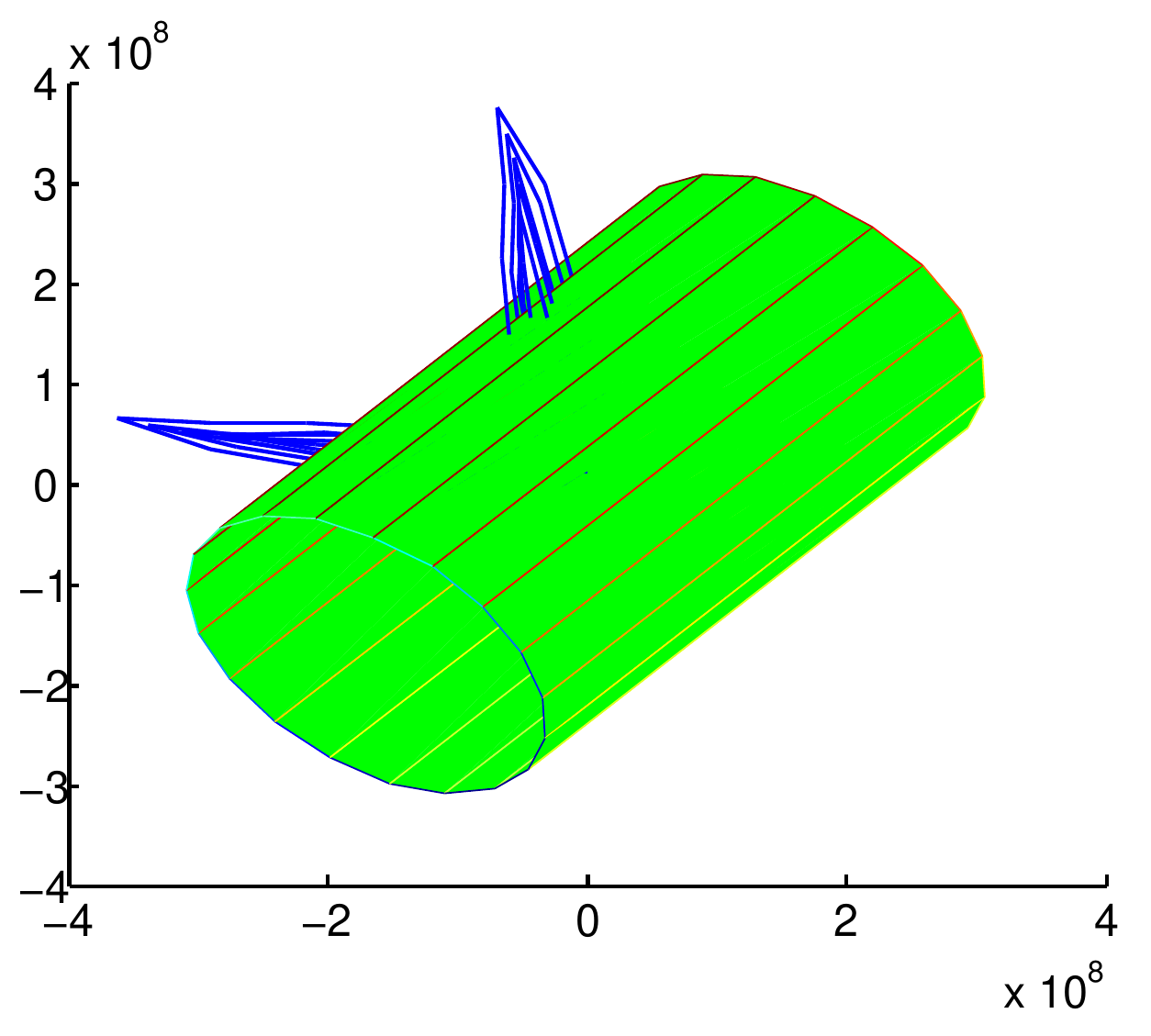}
$\quad$
\includegraphics[width = 2.60in]{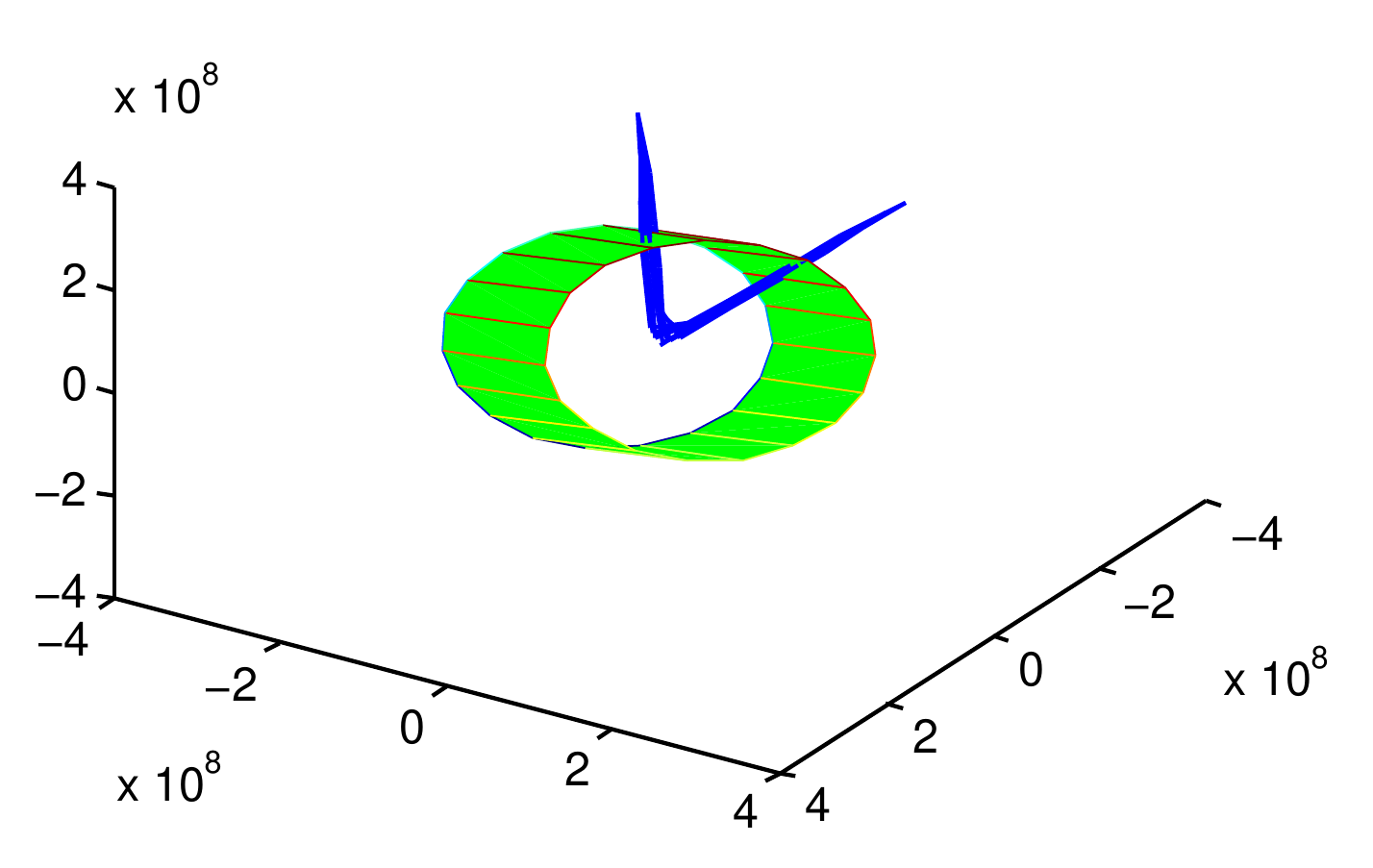}
\caption{\label{fig:VM1} Principal stresses of applied force in 3D considering the von Mises yield criterion- Case 2}
\end{figure}

One reason that can be mentioned is that the process is not always in the states that hardening equations are involved like the elastic states. Therefore less information from the whole simulation can be analyzed for estimating the hardening parameters and updating their parameters' uncertainties. Fig.~\ref{fig:VM} and Fig.~\ref{fig:VM1} prove this fact that since more states are out of the von Mises yield criterion for the second case compared to the first case, in which the hardening equations are involved only in these states, the better identification can be done for the second case, where a gradually varying increasing applied force is considered, for hardening parameters in comparison with the first case where a constant magnitude applied force is employed. In fact, the cyclic applied force in the second case causes the more activation of the desired parameters in the studied set of equations comparing to the first case and accordingly a better determination of the parameters can be carried out for the second case. It should be pointed out that the von Mises yield criterion is illustrated by the green cylinder in the mentioned figures i.e. inside and outside of the cylinder respectively refer to the elasticity and plasticity states, respectively. Also the blue color represents the principal stresses in these figures.


\section {Summary}
Using the Gauss-Markov-Kalman Filter method explained in Section 3 to identify the model parameters of the Choboche model indicates that it is possible to identify the model parameters by employing this method using functional approximation. The parameters are well estimated and the uncertainty of the parameters is reduced while the random variables of the parameters are updated during the process \cite{adeli}. The other conclusion that can be made is that the more information we receive, the better parameter identification we can do using the Gauss-Markov-Kalman filter method. This fact is observed by comparing the posterior probability density functions of hardening parameters, $b_{R}$ and $b_{\vek{\chi}}$ , for case 1 and 2. Therefore in terms of mechanical models, it should be always considered that the applied force should be applied in such a way that all time all equations should be involved. In other words, the applied load path should lead to activation of all uncertain parameters in the set of equations, as here a cyclic gradually varying increasing applied force leads to a better determination of the parameters. Otherwise only the parameters which are in the involved equations are updated.
\\

%
\textbf{\textit{Acknowledgement}} \\
This work is partially supported by the DFG through GRK 2075.
%




\begin{thebibliography}{10}%
%
%
\bibitem{Miller}  A. Miller. An Inelastic Constitutive Model for Monotonic, Cyclic, and Creep Deformation: Part I--Equations Development and Analytical Procedures. J. Eng. Mater. Technol. 98(2), 97--105 (1976).

\bibitem{Krempl}  E. Krempl, J. J. McMahon, and D. Yao. Viscoplasticity Based on Overstress with a Differential
Growth Law for the Equilibrium Stress. Mechanics of Materials 5, 35--48 (1986).

\bibitem{Korhonen1} R. K. Korhonen, M. S. Laasanen, J. Toyras, R. Lappalainen, H. J. Helminen, and J. S. Jurvelin.  Fibril reinforced poroelastic model predicts specifically mechanical behavior of normal, proteoglycan depleted and collagen degraded articular cartilage. J Biomech 36, 1373--1379.

\bibitem{Aubertin}  Michel Aubertin, Denis E. Gill, and Branko Ladanyi. A unified viscoplastic model for the inelastic flow of alkali halides. Mechanics of Materials 11, 63--82 (1991).

\bibitem{Chan}  K. S. Chan, S. R. Bodner, A.F. Fossum, and D.E. Munson. A constitutive model for inelastic flow and damage evolution in solids under triaxial compression. Mechanics of Materials 14, 1--14 (1992).

\bibitem{Chaboche} J. L. Chaboche, and G. Rousselier. On the plastic and viscoplastic constitutive equations - part 1: rules developed with internal variable concept. J. Press. Vessel Technol., 105, 153--158 (1983).

\bibitem{Chaboche1} J. L. Chaboche, and G. Rousselier. On the plastic and viscoplastic constitutive equations - part 2: application of internal variable concepts to the 316 stainless steel. J. Press. Vessel Technol., 105, 159--164 (1983).

\bibitem{KLOSOWSKI} P. K\l osowski, and A. Mleczek. Parameters' Identification of Perzyna and Chaboche
Viscoplastic Models for Aluminum Alloy at Temperature of $120^{\grad}C$. Engng. Trans. 62, 3, 291--305 (2014).

\bibitem{Gong} Y. Gong, C. Hyde, W. Sun, and T. Hyde. Determination of material properties in the Chaboche unified viscoplasticity model. Journal of Materials Design and Applications, 224(1), 19--29 (2010).

\bibitem{Harth} T. Harth, and J\"urgen Lehn. Identification of Material Parameters for Inelastic Constitutive Models Using Stochastic Methods. GAMM-Mitt. 30, No. 2, 409--429 (2007).

\bibitem{Lindholm} K. S. Chan, S. R. Bodner, and U. S. Lindholm. Phenomenological Modelling of Hardening and Thermal Recovery in Metals. Journal of Engineering Materials and Technology 110, 1--8 (1988).

\bibitem{dinkler} J. Velde. 3D Nonlocal Damage Modeling for Steel Structures under Earthquake Loading. Department of Architecture, Civil Engineering and Environmental Sciences
University of Braunschweig - Institute of Technology (2010).

\bibitem{Matthies0} H. G. Matthies. Stochastic finite elements: Computational Approaches to Stochastic Partial Differential Equations. Journal of Journal of Applied Mathematics and Mechanics, volume 88, 849--873 (2008).

\bibitem{Matthies01} H. G. Matthies. Uncertainty Quantification with Stochastic Finite Elements. Encyclopedia of Computational Mechanics, edited by E. Stein, R. de Borst, T. R. J. Hughes, John Wiley and Sons, Chichester, (2007).

\bibitem{Matthies2} H. G. Matthies, E. Zander, B. V. Rosi\'c, and A. Litvinenko. Parameter Estimation via Conditional Expectation: A Bayesian Inversion. Journal of Advanced Modeling and Simulation in Engineering Sciences, 3--24  (2016).

\bibitem{Bojana} B. V. Rosi\'c, A. Litvinenko, O. Pajonk, H. G. Matthies. Sampling-free linear Bayesian update of polynomial chaos representations. Journal of Computational Physics 231(17):5761--5787, (2012).

\bibitem{Matthies3} H. G. Matthies, E. Zander, B. V. Rosi\'c, A. Litvinenko, and Oliver Pajonk. Inverse Problems in a Bayesian Setting. Journal of Computational Methods for Solids and Fluids, volume 41, 245--286 (2016).

\bibitem{Bojana1} B. V. Rosi\'c and H. G. Matthies. Identification of Properties of Stochastic Elastoplastic Systems. Computational Methods in Stochastic Dynamics, 237--253, (2013).

\bibitem{Ghahramani} S. Ghahramani. Fundamentals of probability, with stochastic processes. 3rd ed. New Jersey, USA: Pearson, Prentice Hall (2005).

\bibitem{adeli} E. Adeli, B. V. Rosi\'c, H. G. Matthies and S. Reinst{\"a}dler. Bayesian parameter identification in plasticity. XIV International Conference on Computational Plasticity. Fundamentals and Applications COMPLAS XIV E. O\~nate, D.R.J. Owen, D. Peric and M. Chiumenti (Eds), 247--256, DOI: 10.13140/RG.2.2.22323.89124, (2017).
%
\end{thebibliography}
\end{document}